# Compatibility of Missing Data Handling Methods across the Stages of Producing Clinical Prediction Models

Antonia Tsvetanova[1]; Matthew Sperrin[1]; David A. Jenkins[1]; Niels Peek[2]; Iain Buchan[3]; Stephanie Hyland[4]; Marcus Taylor[5]; Angela Wood [6,7]; Richard D. Riley[8]; Glen P. Martin[1]

1. Division of Informatics, Imaging and Data Science, Faculty of Biology, Medicine and Health, University of Manchester, Manchester Academic Health Science Centre, Manchester, United Kingdom
2. The Healthcare Improvement Studies Institute (THIS Institute), Department of Public Health and Primary Care, University of Cambridge, UK
3. Civic Health Innovation Labs, The University of Liverpool, Liverpool, UK
4. Microsoft Research, Cambridge, UK
5. Department of Cardiothoracic Surgery, Manchester University Hospital NHS foundation Trust, Manchester, UK
6. British Heart Foundation Cardiovascular Epidemiology Unit, Department of Public Health and Primary Care, Victor Phillip Dahdaleh Heart and Lung Research Institute, University of Cambridge, Cambridge, UK
7. Health Data Research UK, London, UK
8. Institute of Applied Health Research, College of Medical and Dental Sciences, University of Birmingham, Birmingham, United Kingdom

**Running Title:** Missing data compatibility across the stages of CPM production

**Funding:** AT was supported by EPSRC and Microsoft Research through an iCASE PhD Scholarship [grant number: 18000110]. MS was supported by the UKRI AI programme, and the Engineering and Physical Sciences Research Council, CHAI - EPSRC AI Hub for Causality in Healthcare AI with Real Data [grant number: EP/Y028856/1]. IB is supported by NIHR as Senior Investigator [NIHR205131]. GPM and RDR are partially supported by an MRC-NIHR Better Methods, Better Research grant [SS-Update: MR/Z503873/1]. RDR is an NIHR Senior Investigator and supported by the NIHR Birmingham Biomedical Research Centre at the University Hospitals Birmingham NHS Foundation Trust and the University of Birmingham. The views expressed are those of the author(s) and not necessarily those of the NHS, the NIHR or the Department of Health and Social Care.

**Competing Interests:** None

**Acknowledgements:** We thank The Steering Committee at the North-West Clinical Outcomes Research Registry for kindly providing the cardiothoracic surgery data.

**Corresponding Author:**
Dr Glen Philip Martin
Senior Lecturer in Health Data Science
Vaughan House, University of Manchester, Manchester, M13 9GB, United Kingdom
Email: glen.martin@manchester.ac.uk

## Abstract

Missing data is a challenge when developing, validating and deploying clinical prediction models (CPMs). Traditionally, decisions concerning missing data handling during CPM development and validation haven't accounted for whether missingness is allowed at deployment. We hypothesised that the missing data approach used during model development should optimise model performance upon deployment, whilst the approach used during model validation should yield unbiased predictive performance estimates upon deployment – we term this 'compatibility'. We aimed to determine which combinations of missing data handling methods across the CPM life-cycle are compatible. We considered scenarios where CPMs are intended to be deployed with missing data allowed or not, and we evaluated the impact of that choice on earlier modelling decisions. Through a simulation study and an empirical analysis of thoracic surgery data, we compared CPMs developed and validated using combinations of complete case analysis, mean imputation, single regression imputation, multiple imputation, and pattern sub-modelling. If planning to deploy a CPM without allowing missing data, then development and validation should use multiple imputation when required. Where missingness is allowed at deployment, the same imputation method must be used during development and validation. Commonly used combinations of missing data handling methods result in biased predictive performance estimates.

# 1 Introduction

Clinical Prediction Models (CPMs) use multiple predictor variables to estimate an individual's risk of developing (prognostic) or currently having (diagnostic) a particular health condition or clinical outcome [1,2]. Widely used examples include QRISK[3], which estimates future cardiovascular disease risk, and EuroSCORE[4], which estimates risk of death shortly after heart surgery. The translation of a CPM into clinical practice typically involves several stages of model development, validation, impact assessment and deployment into practice [2]. Missing data in predictor variables are common at every stage of this CPM life-cycle, which can introduce various biases in the analysis if handled inappropriately[5]. Despite this, there remains limited guidance on how missing data should be handled across the different stages of the CPM life-cycle.

Most methodological research on handling missing data in health research has focused on methods that recover unbiased parameter estimates for inference[6–8]. However, the objectives of prediction model studies differ from those of causal inference or description: the primary goal is to achieve accurate, well-calibrated predictions in new individuals, and this doesn't always require unbiased parameter estimates[9]. As such, existing recommendations of handling missing data for inference do not necessarily apply to prediction modelling studies. For example, multiple imputation (MI) procedures (i.e., chained equations, conditional specification, or other techniques that repeatedly sample missing values conditional on observed data) are widely recommended in medical research because they yield valid parameters estimates and uncertainty quantification when their assumptions hold. However, in the context of CPMs, implementing MI across the entire CPM life-cycle raises practical challenges. For instance, when using MI to handle missing data, the outcome variable must be included in the imputation models[7,10], yet the outcome is unknown at the point of CPM, thus violating the assumption of congeniality between development and deployment. Although methods for applying MI when the outcome variable is systematically missing are emerging[11], many require access to the original development data or computational resources at deployment, limiting their practicality. As a result, alternative approaches – such as single deterministic regression imputation[12] or pattern sub-models[13–15] – have been proposed for prediction settings.

This range of possible imputation methods for CPM research, combined with the current lack of guidance on which methods to use, has created inconsistency in the methods that are being used across the CPM life-cycle[16]. Validation studies frequently apply different missing data handling methods than those used during CPM development, and both can differ from the approach used when the CPM is ultimately deployed[16]. This suggests that decisions about handling missing data during CPM development and validation have historically not been informed by how the CPM is intended to be used in clinical practice. However, Sisk et al. [12] demonstrated that the approach to handling missing predictor variables during development should align with the deployment strategy[12], and Hoogland et al.[17] emphasised that handling missing data during validation should reflect the deployment strategy, ideally using imputation methods that can be transferred alongside the CPM.

Therefore, the decision of how to handle missing data when the CPM is deployed into clinical practice is fundamental in informing the choice of how missing data is handled during CPM development and validation. However, it remains unclear exactly how the intended deployment strategy should inform this choice. In other words, what combinations of missing data approaches used during CPM development and validation are aligned with given deployment strategies? Here, we posit two principles to answer this question:

- First, the method used to handle missing predictor variables during CPM development should minimise model degradation – defined as the difference between a fitted model's performance and the true model's performance – when deploying the CPM under a given missing data strategy; and
- Second, the method used to handle missing predictor variables during CPM validation should provide an unbiased estimate of the CPM's predictive performance when it is deployed under a given missing data strategy.

We define missing data handling methods that achieve both principles as being '*compatible'* across the CPM life-cycle. While using the same missing data handling method across every stage of the life-cycle is clearly compatible (provided the development and validation data are representative of the deployment population), it is unknown what other combinations of missing data handling approach, if any, also satisfy these criteria.

Therefore, the aim of this paper is to identify which combinations of missing data handling methods are compatible and under which missingness assumptions. We distinguish between scenarios where the CPM is deployed allowing for missing data (and if so, how it will be handled), and those in which all predictors must be observed prior to making a prediction. We then examine how this choice impacts missing data handling decisions made during model development and validation, in relation to the above two criteria. Throughout, we assume that the validation data are drawn from the same target population in which the CPM will be deployed, as should be the case in practice[18].

The remainder of the paper is structured as follows. Section 2 describes the missing data handling methods considered and how they are applied during CPM development and deployment. Section 3 presents a simulation study assessing the compatibility of different method combinations under a range of missingness mechanisms. Section 3.7.3 provides an empirical illustration using risk prediction for mortality following lung resection surgery. Section 5 concludes with recommendations for future CPM studies.

# 2 Missing Mechanisms and Missing Data Handling Methods

We begin by describing the different missing data handling approaches we consider in this paper, in the context of how they are (a) used to develop a CPM and (b) incorporated when the CPM is applied to new individuals in a validation dataset or upon model deployment. Throughout, we consider data that are missing completely at random (MCAR), missing at random (MAR), or missing not at random (MNAR), defined depending on whether the probability that a value is missing depends on neither observed nor unobserved data, only observed data, or unobserved data, respectively[19]. Unlike prior work[12,14,17,20], we also allow the missingness mechanism to change across the stages of the CPM life-cycle.

## 2.1 Strategies to Handle Missing Data During Model Development

We denote a development dataset, of $i = 1, \dots, N_{dev}$ individuals, upon which one develops a CPM to estimate the probability of a binary outcome, $Y$ (without loss of generality to other outcome types), conditional on a set of $P$ predictors, denoted $X_1, \dots, X_P$. The development data might contain missing values in the predictors, but we assume the outcome is fully observed; we use $R_{i,p}$ to denote if the $p$-th predictor is missing ($R_{i,p} = 0$) or observed ($R_{i,p} = 1$) for the $i$-th individual and for $p = 1, \dots, P$.

Using these data, we suppose that one wishes to estimate a model of the form:

$$P(y_i = 1|\boldsymbol{x}_i) = g^{-1}(\phi_0, \phi_1, \dots, \phi_P) = \left[1 + \exp\left(-\left(\phi_0 + \sum_{p=1}^{P} \phi_p x_{i,p}\right)\right)\right]^{-1} \quad (1)$$

for unknown parameters, $\phi_1, \dots, \phi_P$ that are estimated within the development data along with the intercept ($\phi_0$). We use a logit-link function in this paper, but note that alternatives are available and that penalty terms might also be added to the likelihood to address overfitting concerns (bias-variance trade-off), such as lasso or ridge penalties. Several strategies are available to fit such a (penalised) regression model in the presence of missing data, which we now outline.

#### *2.1.1 Complete Case Analysis (CCA)*

First, we might choose to fit the CPM (Equation (1)) on the subset of $N_{dev}$ individuals that have complete data on all $P$ predictors - that is, fit the model on all individuals $i$ such that $R_{i,p} = 1 \; \forall \; p \in [1, P]$. CCA is simple to implement but, depending on the proportion of missing data, it can significantly reduce the sample size used to fit the model, leading to a loss of efficiency, increasing the potential of overfitting at model development, and may lead to biased performance estimates at validation under some scenarios[21,22].

#### *2.1.2 Mean/Mode Imputation*

Second, we might use mean/mode imputation, where missing values of continuous predictors are replaced by the mean of the observed values, and missing values of categorical predictors are replaced by the mode of the observed categories (sometimes called "risk factor absent" approach). The mean/mode imputation could be stratified by known variables (e.g., sex, age, ethnicity). Equation (1) is then fit on the resulting complete dataset. This approach is easy to implement but, by imputing all missing values at the mean/mode, variability between individuals is ignored; consequently, standard error estimates of log odds ratios in Equation (1) will be too small and potentially increasing instability of individual's predicted risks[23].

#### *2.1.3 Single (deterministic) Regression Imputation*

Third, we might fit the CPM after applying single regression imputation (RI) to any missing values. That is, we first fit an imputation model to each predictor with missing data, using a suitable regression model with the other predictors as covariates. Each of the obtained regression models are then applied to impute a (single point estimate of the) value of any missing predictor value. This produces a complete dataset, from which Equation (1) can be fitted to develop the CPM. Analytical[10] and empirical[1,12] studies have shown that the outcome should never be included in such deterministic imputation models. Therefore, we only consider RI without the outcome. RI has been shown to perform well for developing CPMs, although can lead to unstable performance estimates compared to multiple imputation [12].

#### *2.1.4 Multiple (stochastic) Imputation*

Fourth, we can use MI by chained equations to develop the CPM. Like Section 2.1.3, a series of imputation models are specified for every predictor with missing data conditional on the other predictors. However, rather than taking just a single point estimate to impute each missing predictor value, one now takes multiple draws from the (posterior distribution of) imputation models, thus creating multiple, slightly different, complete datasets. Equation (1) is then fit in each of the imputed datasets in turn, before the estimates of $\phi_0, \phi_1, \dots, \phi_P$ from each are combined using Rubin's rules to obtain the final CPM [24,25]. Analytical studies have demonstrated that the outcome must be included in such (stochastic) imputation models [10]. However, since the outcome is unavailable during CPM deployment, we also consider MI excluding the outcome; we hypothesised that this approach ensures congeniality (i.e., where the analysis and imputation model are aligned) under scenarios where the CPM will be deployed allowing for missing data, enabling MI models to be applied to new observations upon deployment. We hypothesised that MI including the outcome would be appliable to scenarios where one wishes to develop and validate a CPM that will be deployed requiring all data.

#### *2.1.5 Pattern Sub-Model*

The final approach we consider is the pattern sub-model (PSM). As previously described[13–15], this strategy fits separate CPMs to each combination of non-missing predictors. For ease of explanation, imagine if $P = 2$, then one would fit four CPMs: (i) one with both predictors included as covariates, (ii) one with only $X_1$ included as a covariate, (iii) one with only $X_2$ included as a covariate, and (iv) an intercept-only model. The different models are estimated on the sub-set of individuals who had that corresponding pattern of covariates observed. For example, model (i) would be fit on all individuals with both covariates observed, model (ii) fit on individuals with $X_1$ observed but $X_2$ missing, and so forth. Thus, this creates a set of $2^P$ models to validate (and deploy), wherein one uses the model corresponding to the observed covariate pattern for a new individual. This strategy has been shown to perform well across different missing data mechanisms, but fitting the $2^P$ different sub-models may be computationally challenging and the number of observations for a given pattern of missingness might be too small[13].

### 2.2 Strategies to Handle Missing Data During Model Validation and Deployment

Having developed a CPM using any of the above methods, one now needs to validate it within new individuals of the target population to test its predictive performance. We denote a validation dataset, of size $j = 1, \ldots, N_{val}$, upon which the CPM's predictive performance is evaluated, and which we assume is drawn randomly from the target population where we wish to deploy the CPM[18]. We assume that the validation set contains the same binary outcome $Y$ (fully observed) and the same predictors, $X_1, \ldots, X_P$ (that might contain missing values). We allow for the missingness mechanism between the development and validation datasets to differ but otherwise assume the covariate-outcome joint distributions across the development and validation datasets are the same (i.e., we do not consider data distribution shift other than in missing mechanisms). **Table 1** shows the possible combinations of handling missing data in the validation dataset that we consider in this paper, depending on how missing data were handled during CPM development.

**Table 1**: The combinations of how one can handle missing data in validation, depending on how missing data was handled during CPM development, for each type of imputation approach that we consider in this paper. The aim of this study was to determine which combinations are compatible. The descriptions of each imputation method are exactly as described in Section 2.1.1-2.1.5. CCA = complete case analysis; RI = regression imputation; MI = multiple imputation; PSM = pattern sub-model.

| If a CPM was developed under… | Then it can be validated under… | Explanation |
|---|---|---|
| CCA | CCA | |
| | Mean/Mode imputation<br>MI - fit to the validation data<br>RI - fit to the validation data | If the CPM was developed using CCA, then one can apply CCA to handle missing data in the validation set. Using either RI or MI within the validation set would require new imputation models to be fitted in the validation set, before using them to impute missing data and then apply the CPM to estimate predictive performance. Mean/mode imputation could also be used, taking the summary values as reported from the development set. |
| Mean/Mode Imputation | CCA<br>Mean/Mode imputation<br>MI - fit to the validation data<br>RI - fit to the validation data | If the CPM was developed using mean/mode imputation, then one can use the mean/mode values from the development set to handle missing data in the validation set. Using either RI or MI within the validation set would require new imputation models to be fitted in the validation set, before using them to impute missing data and then apply the CPM to estimate predictive performance. CCA could also be used on the validation set. |
| RI | CCA<br>Mean/Mode imputation<br>MI - fit to the validation data<br>RI - transported from the development data<br>RI - re-fit to the validation data | If the CPM was developed using RI, then missing data can be handled in the validation set using CCA, mean/mode imputation, RI or MI. For RI, one would apply the RI models derived from the development population to impute missing data in the validation set. One could also re-fit the RI models to the validation set (which might be particularly useful if the missingness mechanism changes). Applying MI to the validation set would require new imputation models. |
| MI | CCA<br>Mean/Mode imputation | If the CPM was developed using MI, then missing data can be |
| | MI - transported from the development data<br>MI - re-fit to the validation data<br>RI - fit to the validation data | handled in the validation set using CCA, mean/mode imputation, RI or MI. For MI, one would apply the MI models derived from the development population to impute missing data in the validation set (e.g., using mice::mice.mids() in R). This requires the original MI models/ procedure to be available from the original CPM development study as a 'package' [17]. This is not always possible in practice, so one could also re-fit the MI models to the validation set. Applying RI to the validation set would require new imputation models. |
| PSM | CCA<br>Mean/Mode imputation<br>MI - fit to the validation data<br>RI - fit to the validation data<br>PSM (as estimated from the development data) | If the CPM was developed using the PSMs, then missing data can be handled in the validation set by making predictions from the sub-model corresponding to the missingness pattern for a given individual. Taking the example of $P = 2$ from Section 2.1.5, we would use model (i) to make predictions for individuals in the validation set with fully observed covariates, model (ii) to make predictions for individuals in the validation set with missing $X_2$, and so forth. One could also use CCA, mean/mode imputation, RI (fitting new imputation models to the validation set) or MI (fitting new imputation models to the validation set) to impute missing data, after which the sub-model with all $P$ covariates included (model i in our running example) would be used to make predictions for all individuals. |

# 3 Simulation Study

This section describes a simulation study designed to assess the compatibility of different missing data handling combinations across the stages of CPM development, validation, and deployment. The simulation study was designed according to the ADEMP framework[26,27].

### 3.1 Simulation Aim

The aim of the simulation study was to identify which combinations of missing data handling methods used during CPM development and validation are compatible with a given deployment strategy. To reiterate, compatibility was defined according to two criteria: (1) the missing data handling method used during CPM development should minimise the difference between the CPM's predictive performance and the performance of the 'true' model, when both are deployed under a given missing data strategy; and (2) the missing data handling method used during CPM validation should provide an unbiased estimate of the CPM's performance when it is deployed under a given missing data handling strategy.

### 3.2 Data Generating Mechanism

We simulated an overarching target population of $N = N_{dev} + N_{val}$ observations, comprising two predictor variables, $X_1$ and $X_2$, and a binary outcome, $Y$. The values of $N_{dev}$ and $N_{val}$ were set as 100,000 each, chosen to far exceed minimum sample size requirements for developing[22] and validating[28] a logistic regression CPM across all simulation scenarios we considered. Predictor variable $X_1$ was either continuous or binary depending on the simulation scenario and potentially contained missing data. Predictor variable $X_2$ was always continuous and fully observed. The outcome, $Y$, was also fully observed for all individuals. The predictor variables $X_1$ and $X_2$ were drawn from a bivariate normal distribution, with mean zero, variance 1, and correlation, $\rho$. Across simulation scenarios, $\rho$ took values of 0 and 0.75, which were chosen to reflect situations where the covariates are uncorrelated and quite strongly correlated, respectively. When $X_1$ was binary, it was dichotomised to give a prevalence of 30% (without loss of generality). The outcome was generated according to a logistic regression model:

$$P(Y = 1) = \left[1 + \exp\left(-(\gamma_0 + \gamma_1 X_1 + \gamma_2 X_2)\right)\right]^{-1}$$

where $\gamma_1 \in \{0, 0.5\}$ and $\gamma_2 = 0.5$ across simulation scenarios. These chosen log-odds ratios reflect common values for moderate prognostic factors. For all scenarios, $\gamma_0$ was selected to give an overall outcome prevalence of 20%, reflecting an outcome prevalence that is plausible to see in practice when developing a CPM. Our empirical study in Section 4 considers the case of a rarer outcome. This target population was then split randomly into a development and validation set of sizes $N_{dev} = 100{,}000$ and $N_{val} = 100{,}000$.

Missing data were induced into $X_1$ separately in the development and validation datasets, allowing the missing mechanism to differ between the two stages. Specifically, we generated a binary variable, $R_1$, according to

$$P(R_1 = 1) = \left[1 + \exp\left(-(\beta_0 + \beta_1 X_1 + \beta_2 X_2 + \beta_3 Y)\right)\right]^{-1}$$

such that $X_1$ was missing if $R_1 = 0$. The values of $\beta_1$, $\beta_2$ and $\beta_3$ were selected across simulation scenarios to represent one of six directed acyclic graphs (DAGs; **Figure 1**), each corresponding to different missingness mechanisms[19]. The six DAGs are summarised as follows:

1) DAG (a) represents a situation with no missingness. Here, $P(R_1 = 1) = 1$ for all individuals.

2) DAG (b) represents a MCAR scenario where missingness in $X_1$ does not depend on any observed or missing information. Here, we set $\beta_p = 0$ for $p = 1,2,3$.

3) DAG (c) represents a MAR scenario where missingness in $X_1$ depends only on $X_2$. Here, we set $\beta_1 = \beta_3 = 0$ and $\beta_2 = 0.5$.

4) DAG (d) represents a MCAR scenario where missingness in $X_1$ depends on $X_1$ itself and on $X_2$. We note that complete case analyses can give unbiased parameter estimates under such a mechanism [29]. Here, we set $\beta_3 = 0$ and $\beta_1 = \beta_2 = 0.5$.

5) DAG (e) represents a scenario where missingness in $X_1$ depends on the outcome $Y$ and on $X_2$. This will be MAR when one can condition on $Y$ (e.g., in development where the outcome is known), but MNAR when one cannot condition on $Y$ (e.g., in deployment where the outcome is unknown at the time of making a prediction for a new individual). Here, we set $\beta_1 = 0$ and $\beta_2 = \beta_3 = 0.5$.

6) DAG (f) represents a scenario where missingness in $X_1$ depends on $X_1$ itself, $X_2$ and $Y$. This will either be MNAR depending only on $X_1$ (if one can condition on $Y$) or MNAR depending on both $X_1$ and $Y$ (if one cannot condition on $Y$). Here, we set $\beta_p = 0.5$ for $p = 1,2,3$.

For all DAGs, the values of $\beta_0$ was set to give overall missingness proportions of 10%, 20% and 50% across simulation scenarios. All combinations of development and validation missingness mechanisms were considered. For DAG (a), the versions of the development and validation datasets prior to inducing missing data were stored in each simulation to give "fully observed data" CPMs and performance estimates.

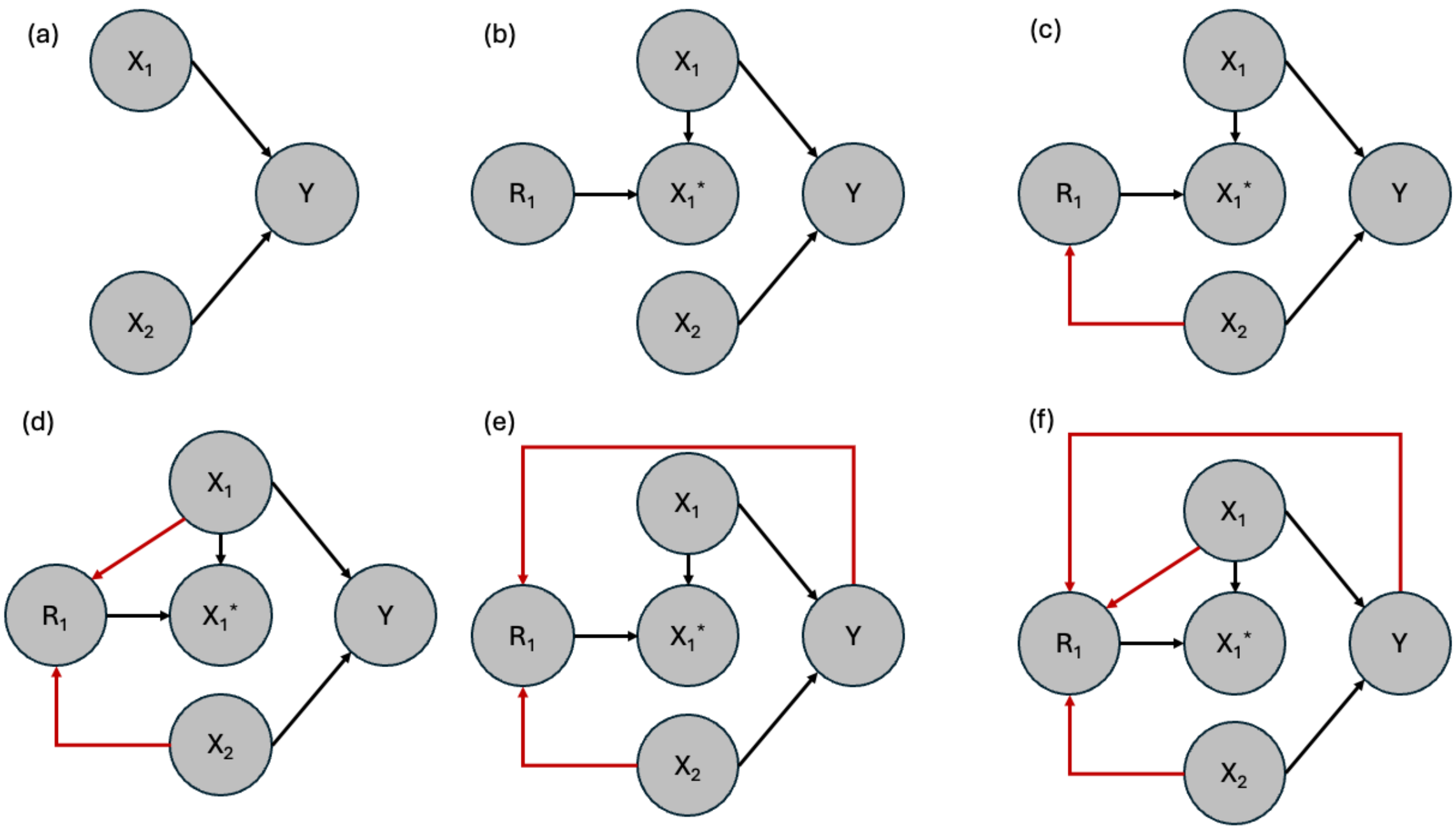

**Figure 1**: Directed acyclic graphs (DAGs) used across the simulation study. $X_1^*$ denotes the observed part of $X_1$.

We considered all combinations of the data-generating parameters (**Table 2**), resulting in 600 scenarios. Each scenario was repeated across 100 iterations, chosen for computational time.

**Table 2**: Summary of the parameters considered in the simulation, and the corresponding values. The simulation was run over all 600 combinations.

| **Parameter Description** | **Parameter Values** |
|---|---|
| $X_1$ data type | Continuous or Categorical |
| Prevalence of $R_1$ | 0.9, 0.8, 0.5 (corresponding to missingness of 10%, 20% and 50%, respectively) |
| Missingness process in development set | DAG (b) – $\beta_1 = \beta_2 = \beta_3 = 0$<br>DAG (c) – $\beta_1 = \beta_3 = 0$ and $\beta_2 = 0.5$<br>DAG (d) – $\beta_3 = 0$ and $\beta_1 = \beta_2 = 0.5$<br>DAG (e) – $\beta_1 = 0$ and $\beta_2 = \beta_3 = 0.5$<br>DAG (f) - $\beta_1 = \beta_2 = \beta_3 = 0.5$ |
| Missingness process in validation set | DAG (b) – $\beta_1 = \beta_2 = \beta_3 = 0$<br>DAG (c) – $\beta_1 = \beta_3 = 0$ and $\beta_2 = 0.5$<br>DAG (d) – $\beta_3 = 0$ and $\beta_1 = \beta_2 = 0.5$<br>DAG (e) – $\beta_1 = 0$ and $\beta_2 = \beta_3 = 0.5$<br>DAG (f) - $\beta_1 = \beta_2 = \beta_3 = 0.5$ |
| Correlation between $X_1$ and $X_2$ ($\rho$) | 0 and 0.75 |
| Prevalence of $Y$ | 20% |
| Effect of $X_1$ on Y ($\gamma_1$) | 0 and 0.5 |
| Effect of $X_2$ on Y ($\gamma_2$) | 0.5 |

### 3.3 Methods

Within each iteration of each simulation scenario, a CPM of the form

$$P(Y = 1) = \left[1 + \exp\left(-(\phi_0 + \phi_1 X_1 + \phi_2 X_2)\right)\right]^{-1} \quad (2)$$

was fit using (unpenalized) maximum likelihood to the fully observed version of the development dataset. Alongside this, missing data in the development dataset was handled using CCA, and imputed using mean/model imputation, RI, MI-no-Y and MI-with-Y, following which, a CPM of the form of Equation (2) was fit as described in Section 2. MI was implemented using chained equations (MICE). The PSMs were fit in the development set using the methods of Fletcher Mercaldo and Blume [13]. For RI and MI, linear ($X_1$ continuous) or logistic ($X_1$ binary) regression was used as the underlying imputation models, and for MI we generated 5 imputed datasets chosen for computational reasons. For both RI and MI, the imputation models included $X_2$ (and possibly also $Y$ for MI-Y).

We then imputed the validation dataset under CCA, RI, MI-no-Y, MI-with-Y, and PSM. For RI and MI, we considered situations where the imputation models were transported from the development dataset, and where the imputation models were re-fit to the validation dataset. Every developed CPM was then applied to make predictions in every imputed validation dataset (across the possible combinations as outlined in **Table 1**). Every CPM was also applied to the fully observed version of the validation dataset.

### 3.4 Estimand (Deployment Strategies)

Our estimand is the predictive performance of a CPM when *deployed* under a specific missing data handling strategy. We considered five deployment strategies:

- E-all: deployment requiring fully observed predictor variables
- E-mean: deployment using mean/mode imputation

- E-RI: deployment using RI
- E-MI: deployment using MI-no-Y
- E-PSM: deployment using the PSM approach

These were chosen to match deployment strategies that are commonly seen in practice[16].

### 3.5 Performance Measures

We describe predictive performance in terms of calibration (agreement between observed and predicted risk), discrimination (ability of the model to assign higher predicted risks to those that have the outcome) and overall model accuracy. We quantify these using the calibration intercept, calibration slope, area under the receiver operating characteristic curve (AUC) and Brier Score.

For each estimand (deployment strategy) specified in Section 3.4, we calculated:

(a) The 'true predictive performance': defined as the performance of the data-generating model evaluated in the validation dataset where missing data was handled using the approach corresponding to the given estimand. For example, the true E-all performance was the performance of the data generating model within the fully observed validation dataset, the true E-RI performance was the performance of the data generating model within the validation dataset imputed using RI, and so forth.

(b) The 'deployment predictive performance' of a given CPM: defined as the performance of a given CPM estimated within the validation dataset where missing data was handled using the approach corresponding to the given estimand. For example, the E-all deployment performance of a given CPM is the performance of that CPM estimated in the fully observed validation dataset, the E-RI deployment performance of a given CPM is the performance of that CPM estimated in the validation dataset imputed using RI, and so forth. This definition reflects our assumption throughout that the validation data are drawn from the same target population in which the CPM would be deployed in practice, as should be the case[18].

These two quantities then allowed us to calculate our two primary performance indicators, which together form our definition of compatibility (see Section 1). First, model degradation of a given CPM, for each estimand. Model degradation was defined as the difference between the CPM's deployment performance for a given estimand, and the true performance for that estimand. For example, the model degradation of a CPM developed using CCA for the E-RI estimand was the performance of that CPM estimated within the validation set imputed using RI, minus the true E-RI predictive performance. Second, validation bias of a given CPM, for each estimand. Validation bias was defined as the difference between a CPM's deployment performance and its validated performance under all alternative missing data handling strategies. For example, validation bias for a CPM developed using RI for the E-MI estimand was the performance of that CPM estimated in the validation dataset imputed using every missing data handling method, minus the performance of that CPM estimated in the validation dataset imputed using MI. Thus, validation bias will, by definition, be zero when the missing data handling methods are the same between validation and deployment; this matches the criteria we use to define compatibility.

## 3.6 Software

The simulation was implemented using R version 4.5.0[30], along with the tidyverse, mice, pROC and ggh4x packages. The full simulation code and results can be found on the following GitHub page: https://github.com/antoniatsv/R-Compatibility-Sim-Study.

## 3.7 Simulation Study Results

For ease of illustration, we present results for scenarios where $X_1$ is continuous, contains 50% missing data, $\rho = 0.75$, and where $\gamma_1 = \gamma_2 = 0.5$. Results were quantitively similar in all other scenarios (the full results are available on the GitHub repository). We now summarise the results for each of our five estimands, in turn.

### *3.7.1 E-all: deploy under all data required*

If targeting an all data required deployment strategy (E-all) and if the missingness mechanism was MCAR (DAG(b)), then developing the CPM under CCA, MI-with-Y, PSM or developing on fully observed data, resulted in minimal model degradation (**Supplementary Figure 1**). Similarly, imputing the validation set using CCA or MI-with-Y, applying the PSM, or using a fully observed validation data, resulted in unbiased performance estimates of E-all (**Supplementary Figure 2**).

When the missingness mechanism was MAR (DAG(c)), then developing the CPM under CCA, MI-with-Y, PSM or developing on fully observed data, resulted in minimal model degradation for targeting an E-all deployment strategy (**Figure 2**). However, validating under CCA when MAR resulted in a small degree of bias for targeting an E-all deployment strategy (**Figure 3**); there was also a small degree of bias in the AUC estimate when using the PSM to validate targeting E-all when the correlation between predictors was reduced to zero. In contrast, imputing the validation set using MI-with-Y or using a fully observed validation data, resulted in unbiased performance estimates of E-all (**Figure 3**).

When missingness in $X_1$ depended on $X_1$ and on $X_2$ (DAG(d)), then we found degradation in the calibration of the CPMs developed using MI (both with and without Y), RI and mean imputation (**Supplementary Figure 3**). Developing a CPM using the PSM or CCA resulted in minimal model degradation under DAG(d) missingness mechanism for E-all deployment. However, under this missingness mechanism, none of the missing data handling methods in the validation set resulted in unbiased performance estimates of E-all (except validating under a fully observed dataset, by definition), regardless of how the CPM was developed (**Supplementary Figure 4**).

For cases where missingness depended on the outcome (DAG(e)), then only developing a model using MI-with-Y or developing on fully observed data resulted in minimal model degradation upon deploying under E-all (**Supplementary Figure 5**). Similarly, only imputing the validation data using MI-with-Y or validating on fully observed data gave unbiased estimates of E-all deployment performance, regardless of how the CPM was developed (**Supplementary Figure 6**). However, when missingness depended on $X_1$ and $Y$ (DAG(f)), then developing a CPM under any imputation approach resulted in model degradation when deployed under E-all, especially in terms of calibration (**Figure 4**). Similarly, under this missingness mechanism, all approaches to handle missing data in validation resulted in biased performance of E-all deployment, regardless of how the CPM was developed (**Figure 5**).

We found similar results when the missingness mechanism changed between development and validation. For example, if the development dataset was missing under DAG (c) and the validation dataset was missing under DAG(f), then unbiased performance results of E-all were only obtained by imputing the validation set using MI-with-Y. Refitting the imputation models within the validation set offered some protection, but did not eliminate biases.

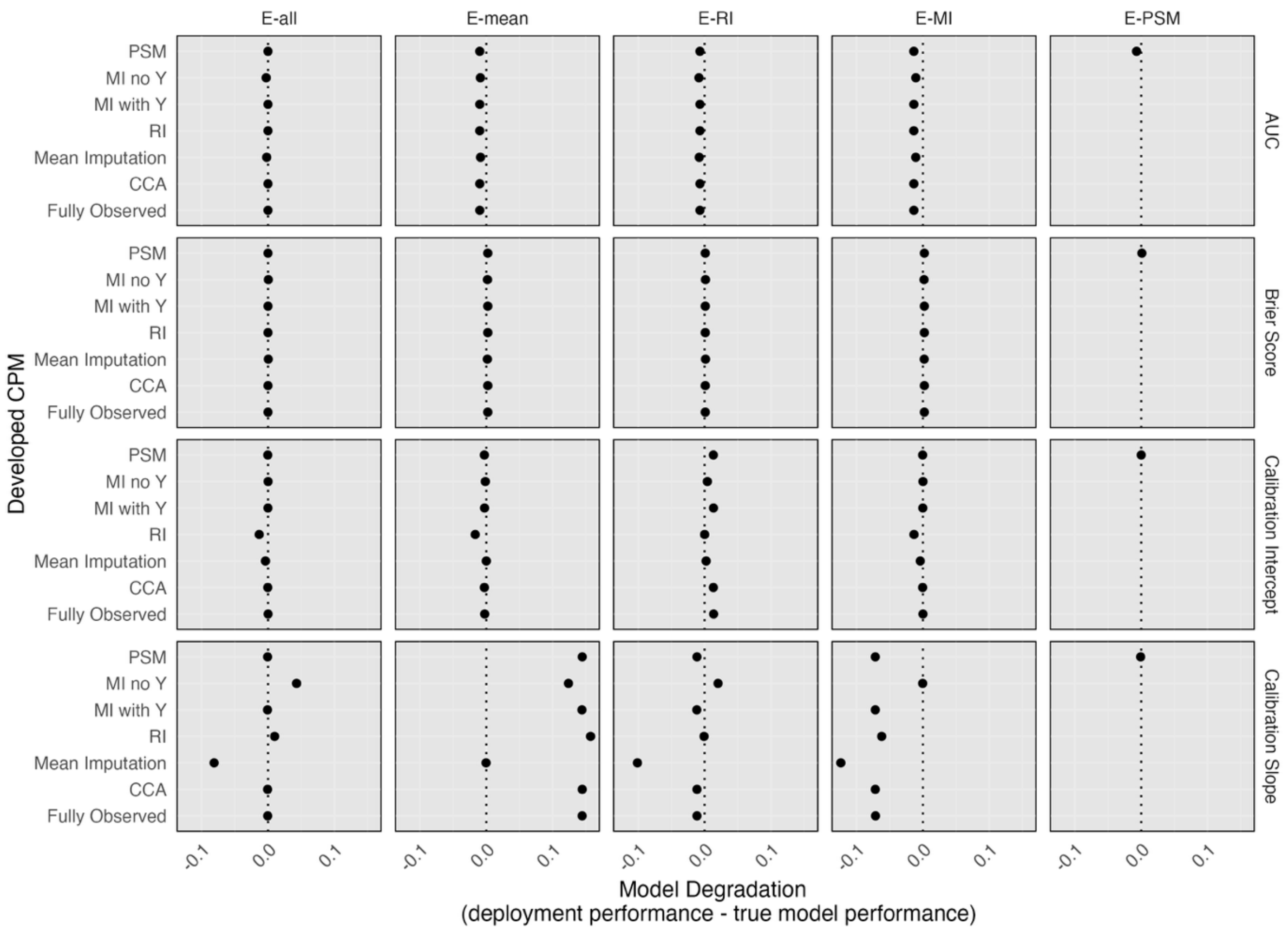

**Figure 2:** Model Degradation of each developed CPM for each estimand, for scenarios were $X_1$ is continuous, contains 50% missing data, where $\gamma_1 = \gamma_2 = 0.5$, where the correlation between $X_1$ and $X_2$ ($\rho$) was 0.75, and under a consistent DAG(c) missingness mechanisms in development and validation sets. Model degradation was defined as the difference between the CPM's deployment performance for a given estimand, and the true performance for that estimand.

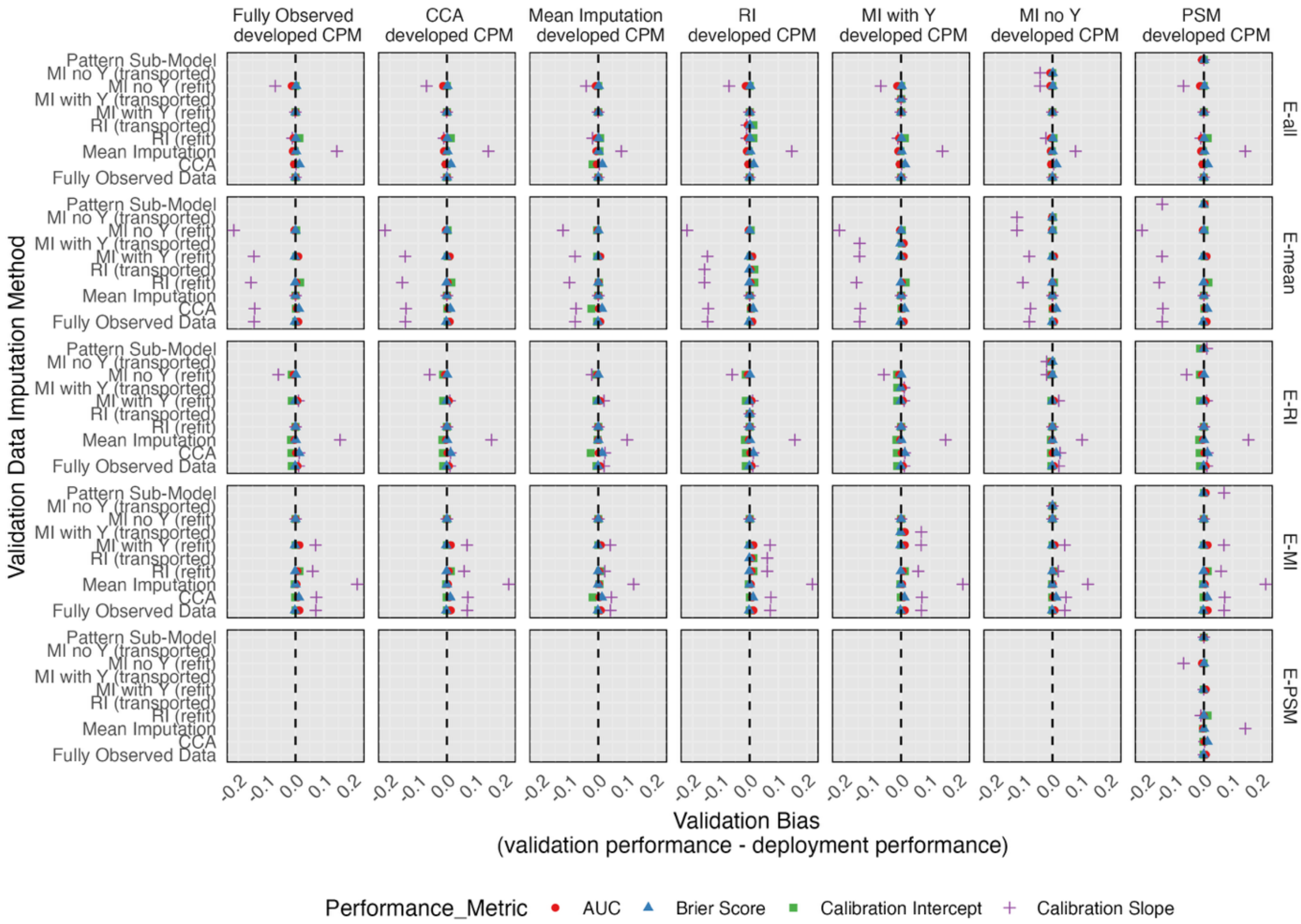

**Figure 3:** Validation Bias of each developed CPM for each estimand, for scenarios were $X_1$ is continuous, contains 50% missing data, where $\gamma_1 = \gamma_2 = 0.5$, where the correlation between $X_1$ and $X_2$ ($\rho$) was 0.75, and under a consistent DAG(c) missingness mechanisms in development and validation sets. Validation bias was defined as the difference between a CPM's deployment performance and its validated performance under all alternative missing data handling strategies.

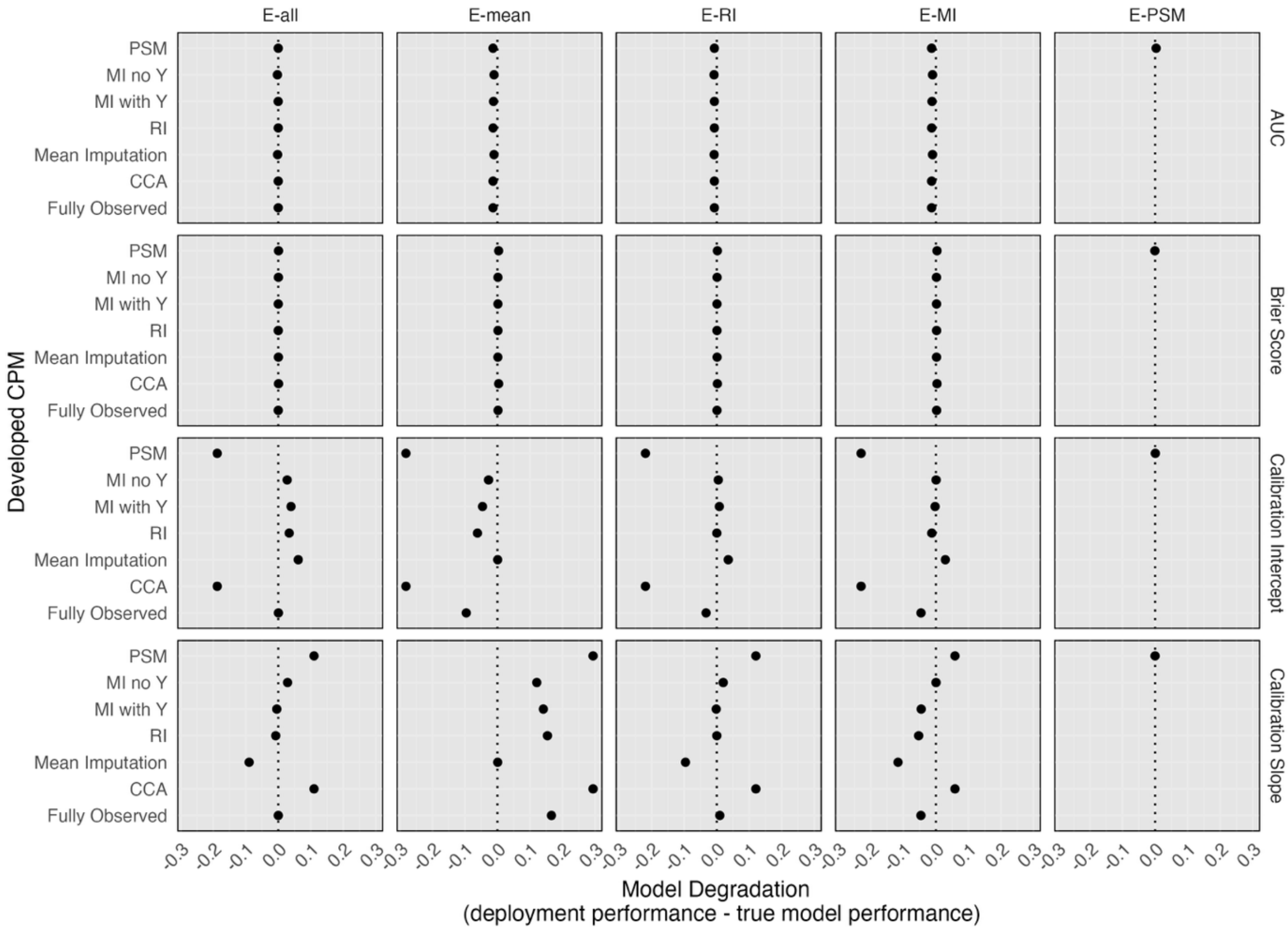

**Figure 4:** Model Degradation of each developed CPM for each estimand, for scenarios were $X_1$ is continuous, contains 50% missing data, where $\gamma_1 = \gamma_2 = 0.5$, where the correlation between $X_1$ and $X_2$ ($\rho$) was 0.75, and under a consistent DAG(f) missingness mechanisms in development and validation sets. Model degradation was defined as the difference between the CPM's deployment performance for a given estimand, and the true performance for that estimand.

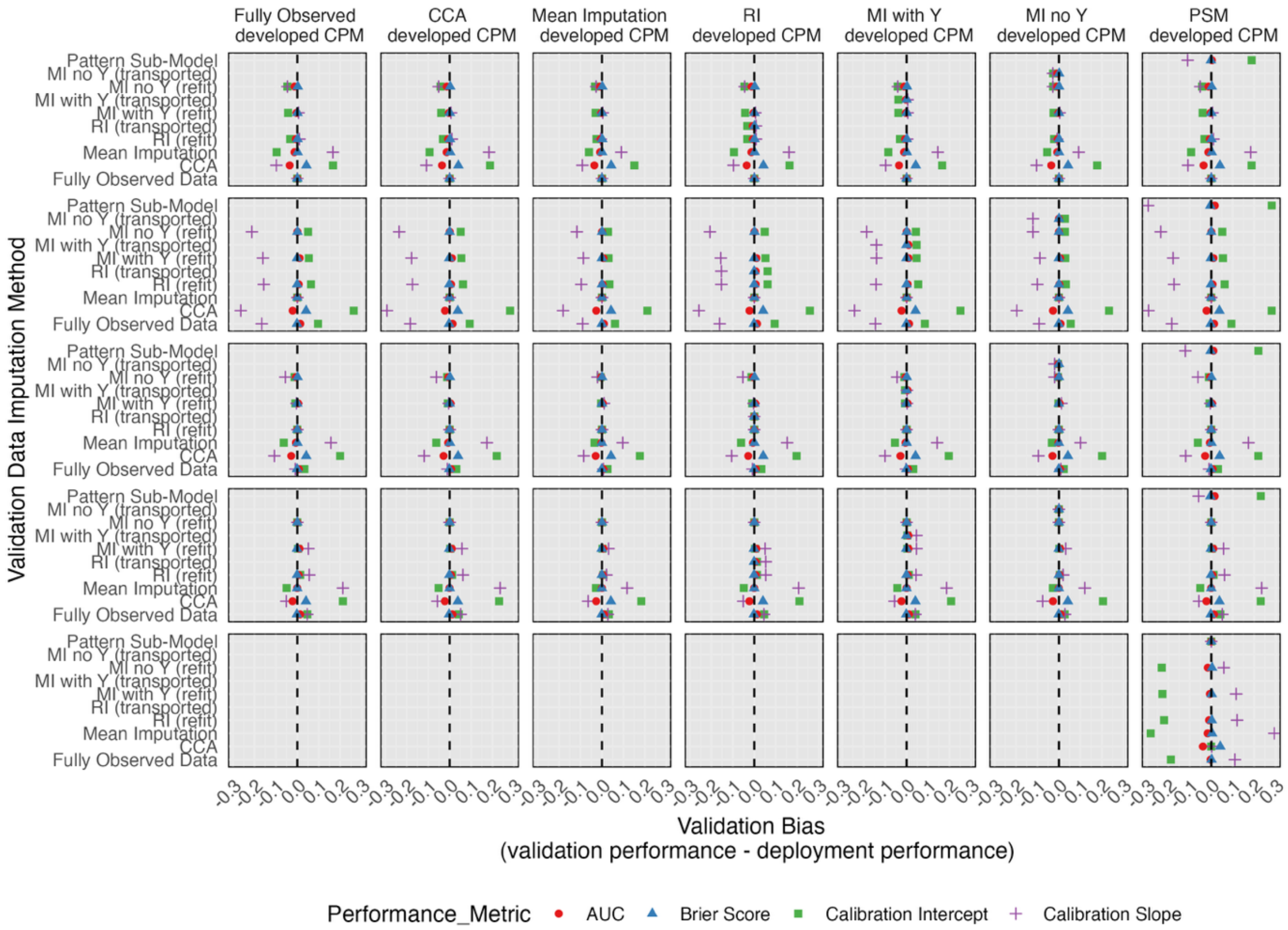

**Figure 5:** Validation Bias of each developed CPM for each estimand, for scenarios were $X_1$ is continuous, contains 50% missing data, where $\gamma_1 = \gamma_2 = 0.5$, where the correlation between $X_1$ and $X_2$ ($\rho$) was 0.75, and under a consistent DAG(f) missingness mechanisms in development and validation sets. Validation bias was defined as the difference between a CPM's deployment performance and its validated performance under all alternative missing data handling strategies.

### *3.7.2 E-mean: deploy under a mean/model imputation strategy*

Developing the CPM using any imputation method other than mean/mode imputation resulted in a degradation of calibration upon deploying those models under E-mean, across all missingness mechanisms (**Figures 2** and **4**, **Supplementary Figures 1, 3** and **5**). Likewise, imputing the validation set using any method other than mean/mode imputation also resulted in biased estimates of E-mean deployment performance, regardless of how the CPM was developed (**Figures 3** and **5**, **Supplementary Figures 2, 4** and **6**).

### *3.7.3 E-RI: deploy under a RI imputation strategy*

When data were MCAR (DAG (b)) or MAR (DAG (c)), then there was model degradation in calibration of CPMs developed using any method other than RI, when deploying under E-RI strategy (**Figure 2** and **Supplementary Figure 1**). Likewise, imputing the validation data using mean/mode imputation or CCA resulted in biased estimates of E-RI performance, regardless of how the CPM was developed (**Figure 3** and **Supplementary Figure 2**). Imputing the validation dataset using MI (with or without the outcome) also resulted in biased E-RI deployment performance, albeit the level of bias for MI-with-Y was relatively small. Despite being intuitively similar to RI, imputing validation data using MI-no-Y led to biased estimates of E-RI because (as shown analytically by McGowan et al.[10]) failing to include the outcome in stochastic imputation biases the estimates of the association between predictor variables and outcome; this becomes apparent with bias in the calibration slope (**Figure 3** and **Supplementary Figure 2**). Validating any of the CPMs using fully observed data was slightly biased for estimating E-RI deployment performance. Validating a PSM developed CPM using the PSM approach in the validation dataset performed well at estimating E-RI deployment performance.

When the data were missing under DAGs (d)-(f), we found that developing the CPM using any method other than RI resulted in degradation of E-RI performance (**Figures 4**, **Supplementary Figures 3** and **5**). The level of degradation was relatively small for MI-with-Y. In terms of validation bias, imputing the validation set using any method other than RI was biased of E-RI performance under these missingness mechanisms, regardless of how the CPM was developed (**Figures 5**, **Supplementary Figures 4** and **6**). Similar observations were made when missingness mechanism changed between development and validation.

#### *3.7.4 E-MI: deploy under a MI-no-Y imputation strategy*

When deploying the model under a MI-no-Y imputation strategy (E-MI), then there was degradation in model performance – especially calibration slope – across all methods of developing the CPM, except using MI-no-Y (**Figures 2** and **4**, **Supplementary Figures 1, 3** and **5**). Similarly, all approaches to impute missing validation data (other than MI-no-Y, by definition) resulted in biased estimates of E-MI deployment performance, regardless of how the CPM was developed (**Figures 3** and **5**, **Supplementary Figures 2, 4** and **6**).

#### *3.7.5 E-PSM: deploy under a PSM strategy*

If aiming to deploy the model under a PSM strategy (E-PSM), then one needs to develop the CPM under that approach (by definition). When deploying the model under a PSM strategy, then the only unbiased method was to also impute the validation dataset under the same imputation strategy. Imputing the validation dataset using MI, RI, mean/mode or CCA all resulted in biased estimates of E-PSM deployment performance (**Figures 3** and **5**, **Supplementary Figures 2, 4** and **6**).

# 4 Empirical Example

We now describe the methods and results of our empirical study of risk prediction for mortality following lung resection surgery. The aim of this analysis was to evaluate how different missing data handling strategies perform in real-world data, under conditions broadly similar to those explored in the simulation study. Unlike the simulations, this empirical example considers a rarer outcome, allowing us to assess whether outcome prevalence influences our conclusions.

## 4.1 Data Source, Study Population and Predictor Variables

We used data form the North-West Clinical Outcomes Research Registry (NCORR) and included patients who underwent lung resection between January 2012 and December 2018 at either Manchester University NHS Foundation Trust or Liverpool Heart & Chest Hospital. We excluded those undergoing alternative surgical procedures. The registry captures pre-operative, peri-operative, and post-operative data, with post-operative outcomes added upon patient discharge. Variables with more than 20% missingness were removed prior to us receiving the data. Data were anonymised before analysis, and individual patient consent was not required under the registry's ethical approval. The study was approved by the NCORR steering committee.

We aimed to develop a CPM to predict death within 90 days of surgery, and considered the same predictor variables and functional form as the existing RESECT-90 CPM[31]. This CPM was also developed using the NCORR dataset to predict 90-day mortality post-surgery. Specifically, the predictors were age (continuous), sex (categorical), performance status (categorical), diffusion capacity of the lung for carbon monoxide (continuous), body mass index (continuous), creatinine (continuous), anaemia (categorical), arrhythmia (categorical), right-sided resection (categorical), number of resected bronchopulmonary segments (continuous), surgery via thoracotomy (categorical) and confirmed/suspected diagnosis of malignancy (categorical).

### 4.2 Model Development and Validation

We took bootstrap samples (with replacement) from the NCORR data, creating 100 'development datasets'. Within each development dataset, we applied the same missing data handling methods as described in Section 2. For each missing data handling method, we then developed a logistic regression CPM to predict the primary outcome.

After development, each CPM (along with any associated imputation model) was applied to the original NCORR dataset, which acted as the validation dataset. Like in the simulation study, missing data in the validation dataset were handled using the missing data strategies appropriate for a given CPM (**Table 1**). This allowed us to evaluate all relevant combinations of development and validation approaches. Predictive performance was estimated for each CPM in each imputed validation dataset, and this process was repeated across all 100 development bootstrap samples. The performance estimates of each CPM in each imputed validation set were then summarised across the bootstrap replications.

As per the simulation study, we evaluated predictive performance with respect to different deployment strategies (Section 3.4). For each deployment strategy, we identified the validation results that corresponded to the intended deployment missingness approach and treated those estimates as the deployment performance. We then calculated differences between this deployment performance and the performance obtained under alternative ways of imputing the validation dataset. These differences mimic the 'validation bias' metric used in the simulation study. Because fully observed NCORR data were not available, we were unable to evaluate the estimand corresponding to deployment requiring complete data (E-all).

### 4.3 Empirical Study Results

After applying our inclusion and exclusion criteria, the NCORR dataset included 6600 patients. The mean age was 66.8 years, and 50% of the cohort were female (**Table 3**). The following variables contained missing data: sex (6/6600), performance status (103/6600), diffusion capacity of the lung for carbon monoxide (1208/6600), body mass index (569/6600) and creatinine (606/6600). The primary outcome occurred in 204 patients (3.09%).

**Table 3:** Baseline summary of the NCORR data included in our empirical study, both overall and by primary outcome status.

| **Characteristic** | **Overall**, N = 6,600[1] | **Alive at 90 days**, N = 6,396[1] | **Dead at 90 days**, N = 204[1] |
|---|---|---|---|
| Age (years) | 68 (61, 74) | 68 (61, 74) | 72 (67, 77) |
| Sex | | | |
| Female | 3,313 (50%) | 3,240 (51%) | 73 (36%) |
| Male | 3,281 (50%) | 3,151 (49%) | 130 (64%) |
| Unknown | 6 | 5 | 1 |
| Performance Status | | | |
| 0 | 2,200 (34%) | 2,152 (34%) | 48 (24%) |
| 1 | 3,898 (60%) | 3,768 (60%) | 130 (65%) |
| 2 | 345 (5.3%) | 325 (5.2%) | 20 (10%) |
| 3 | 54 (0.8%) | 53 (0.8%) | 1 (0.5%) |
| Unknown | 103 | 98 | 5 |
| Diffusion capacity of the lung for carbon monoxide | 71 (59, 85) | 72 (59, 85) | 60 (49, 73) |
| Unknown | 1,208 | 1,168 | 40 |
| Body Mass Index | 26.4 (23.1, 30.0) | 26.5 (23.2, 30.0) | 24.6 (22.0, 28.0) |
| Unknown | 569 | 553 | 16 |
| Creatinine (umolL) | 74 (65, 87) | 74 (65, 87) | 76 (67, 94) |
| Unknown | 606 | 585 | 21 |
| Anaemia | 1,511 (23%) | 1,434 (22%) | 77 (38%) |
| Arrhythmia | 423 (6.4%) | 398 (6.2%) | 25 (12%) |
| Right-side resection | 4,010 (61%) | 3,875 (61%) | 135 (66%) |
| Number of Resected Segments | 3.00 (2.00, 5.00) | 3.00 (2.00, 5.00) | 4.00 (3.00, 5.00) |
| Surgery via thoracotomy | 4,744 (72%) | 4,569 (71%) | 175 (86%) |
| Diagnosis of malignancy | 6,030 (91%) | 5,831 (91%) | 199 (98%) |
| [1] Median (IQR); n (%) | | | |

**Figure 6** presents the difference in estimated validation performance under each imputation approach and the deployment performance of each deployment strategy for every developed CPM. When targeting E-mean deployment, we found that if the CPM was developed using mean/mode imputation, then imputing the validation set under any other approach (except CCA) resulted in small differences to the E-mean deployment performance. For other developed CPMs, then only imputing the validation set using mean/model imputation gave close estimation to the E-mean deployment performance.

We found that imputing the validation data under either MI or RI generally gave close agreement with E-RI and E-MI deployment performance, regardless of how the CPM was developed (**Figure 6**). For E-RI and E-MI deployment strategies, imputing the validation dataset using PSM, mean/mode or CCA resulted in quite large differences between estimated performance and deployment performance.

Finally, if targeting deploying the CPM under PSM, then the CPM needs to be developed using the PSM approach (by definition), and therein we found that validating the model using any other missing data handling method gave large differences to the E-PSM performance (bottom-right panel of **Figure 6**).

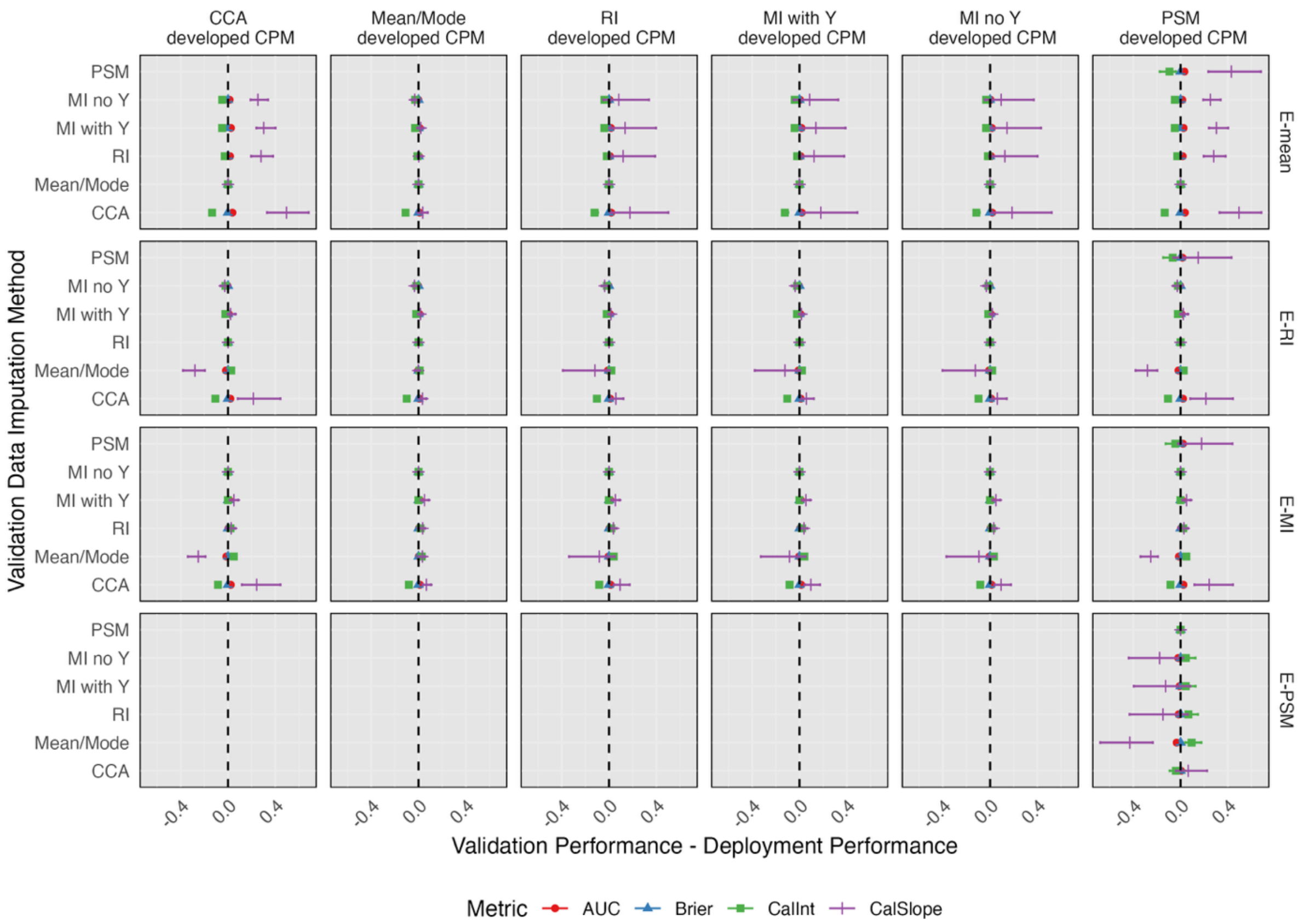

**Figure 6:** Bias in predictive performance results for each developed CPM (panel rows) upon validation using different missing data handling methods (y-axis), when targeting deployment under a given missing data handling strategy (panel columns).

## 5 Discussion

In this study, we investigated how missing predictor data should be handled across the stages of CPM development, validation and deployment. Our aim was to identify missing data handling strategies that minimise model degradation between development and deployment, and strategies during model validation that provide an unbiased estimate of the CPM's predictive performance at deployment. We considered a range of missing data handling methods, including those that are most commonly used across the stages of CPM production in practice[16] and emerging approaches based on pattern sub-models[13–15]. Using a large simulation study and an empirical example, we demonstrated that the intended handling of missing data at deployment has important implications for how missing data should be handled earlier in the CPM life-cycle. Our findings show that many commonly used combinations of missing data handling methods across development, validation, and deployment lead to biased or sub-optimal predictive performance. Compatibility can be achieved when decisions at each stage are aligned with the intended deployment strategy.

### 5.1 Deploying CPMs When Missing Data Are Not Allowed

When a CPM is intended to be deployed only in individuals with fully observed predictor data, we found that the model should be developed and validated using either fully observed data or multiple imputation that includes the outcome (MI-with-Y). These approaches resulted in minimal model degradation and unbiased estimates of deployment performance across most missingness mechanisms. This finding aligns with previous work showing that MI-with-Y is required to obtain valid parameters estimates and predictions when targeting complete-data deployment scenarios[12]. If a CPM has been developed using MI-with-Y, then validation can be conducted either by transporting the imputation models to the validation dataset[17] or by refitting the imputation models within the validation data. The latter may be preferable when the original imputation models are unavailable, when the original CPM was not developed using MI, or when the missingness mechanism is suspected to differ between stages of CPM production. For example, the missingness mechanism might change between development and deployment if, after developing a CPM, an important predictor – previously overlooked due to high missingness (perhaps difficult to observe) – is identified, prompting increased efforts to collect it during implementation of the CPM. Regardless, including the outcome in the imputation models is a requirement, as shown previously[10].

However, we also observed scenarios – particularly under extreme forms of MNAR – where MI-with-Y resulted in biased estimates of E-all deployment performance. In such cases, alternative approaches, such as the inclusion of missingness indicators, may offer partial protection, as previously reported[12].

### 5.2 Deploying CPMs When Missing Data Are Allowed

When a CPM is intended to be deployed allowing for missing predictor values, our results consistently showed that the same missing data handling method should be used during development, validation and deployment, to minimise degradation and gave accurate validation performance estimates. While some alternative combinations appeared compatible under some specific scenarios (e.g., PSM was compatible with E-RI under MCAR and MAR), these observations did not hold generally.

Accordingly, we recommend that researchers first consider how missing data are likely to arise in the deployment setting, drawing on clinical context and knowledge of the data collection processes, to inform how the model will be deployed. Once a deployment strategy has been selected, decisions regarding missing data handling at development and validation should be made to match that strategy. Working backwards from deployment in this way provides the safest and most robust approach.

Our findings demonstrate that several combinations of methods commonly used in practice lead to biased and sub-optimal models[16]. For example, validating a CPM using MI and then deploying it using mean/mode imputation – a combination frequently observed in published studies[16] – resulted in biased performance estimates. This is unsurprising since MI is designed to provide unbiased estimates as if the data had not been missing[6,24], which is not the case for mean/mode imputations.

An important finding was that validation bias depended primarily on how missing data were handled during validation, rather than on how the CPM was originally developed. Consequently, when externally validating an existing CPM, researchers should handle missing data in the validation dataset using the same strategy that is planned for deployment, even if this differs from the method used during development of the original model.

## 5.3 Implications for CPM Production

Our recommendation to align missing data handling across development, validation, and deployment reinforces previous recommendations that we should use missing data handling methods that can be readily applied to new patients[17]. For example, deploying the CPM under an MI approach would be challenging in practice (e.g. where the original development data, imputation models, or sufficient computational power are not available), although adaptions to the standard MI procedures have been proposed to facilitate this[17,20]. As such, if planning to deploy a CPM where missing data is allowed, we recommend developing and validating the model using either regression imputation or pattern sub-models, as strategies that are both easy to implement in deployment and are easy to report for application in new individuals without needing access to the raw development/ validation data.

Together, this study and prior work[12,17], emphasises the need for CPM developers to consider deployment strategies at the outset, as this choice has ramifications for how missing values are handled during CPM development and validation. Specifically, distinguishing between a model that is optimised for use for individuals with complete data, and a model which provides predictions for people with incomplete data. This decision should be informed by the clinical context and made in consultation with end-users, prior to model development. Of course, early engagement with stakeholders has additional benefits beyond missing data handling, such as facilitating alignment between the CPM, clinical workflows, and implementation constraints, to increase the likelihood of translation into practice[32]. In some clinical settings, end-user engagement may reveal that a CPM should support both complete-data and incomplete-data deployment cases. For example, QRISK[3] was developed targeting an E-all scenario (i.e., developed using MI-with-Y), but is deployed using mean/mode imputation to allow predictions in individual's with missing predictors. Our findings suggest that such situations would ideally require separate models, each developed and validated specifically for their intended deployment scenario; this is not commonly done in practice.

## 5.4 Limitations

Several limitations warrant discussion. First, the simulation study focused on CPMs with two predictors, only one of which contained missing data. While this setup can be viewed as the two covariates as representing summaries of multiple missing and observed predictors, future work should explore higher-dimensional predictor spaces. Second, we considered only regression-based models for both the CPM itself and the underlying imputation models; extending this work to machine learning approaches would be valuable (e.g., Random Forest imputation[33]). Third, we defined validation bias to be the difference in performance estimated in validation data that matched the missing data handling of each estimand and the performance of other missing data handling methods. This means that using the same missing data handling method in validation as that defined in the target estimand was unbiased by design. We believe this is an appropriate choice, but we acknowledge there are potentially other definitions of compatibility that we didn't consider. Related to this, we did not deploy any of the CPMs, but instead we took the performance in validation to represent the performance that would be observed if the model were deployed. This assumes that the validation data sufficiently represents the target population where the model will be deployed, which should be true in practice[18]. Fourth, the sample sizes used to develop and validate the CPMs were large[22,28], and we did not explore the potential interplay between minimum required sample sizes and strategies for handling missing data, which could be considered for future work. Finally, within the empirical study, the data were cleaned prior to us receiving the data such that variables with more than 20% missingness were removed, which may have attenuated differences between missing data handling strategies.

### 5.5 Conclusion

The choice of how to handle missing data during CPM development and validation should be driven by how missing data will be handled when the CPM is deployed, alongside prior recommendations to always careful consider the expected missingness mechanism in deciding how to handle missing data. If the CPM is intended to be deployed without allowing for missing data, then the model should be developed and validated in either fully complete data or data imputed using multiple imputations (including the outcome in the imputation models). Conversely, if the CPM is intended to be deployed allowing for missing data, then one should use an identical approach to handle missing data during development and validation to the planned missing data strategy during CPM deployment. Commonly used combinations of missing data handling methods across the stages of CPM production result in biased predictive performance estimates.

<u>**Supplementary Material: Compatibility of Missing Data Handling Methods across the Clinical Prediction Model Pipeline**</u>

# 7 Supplementary Figures

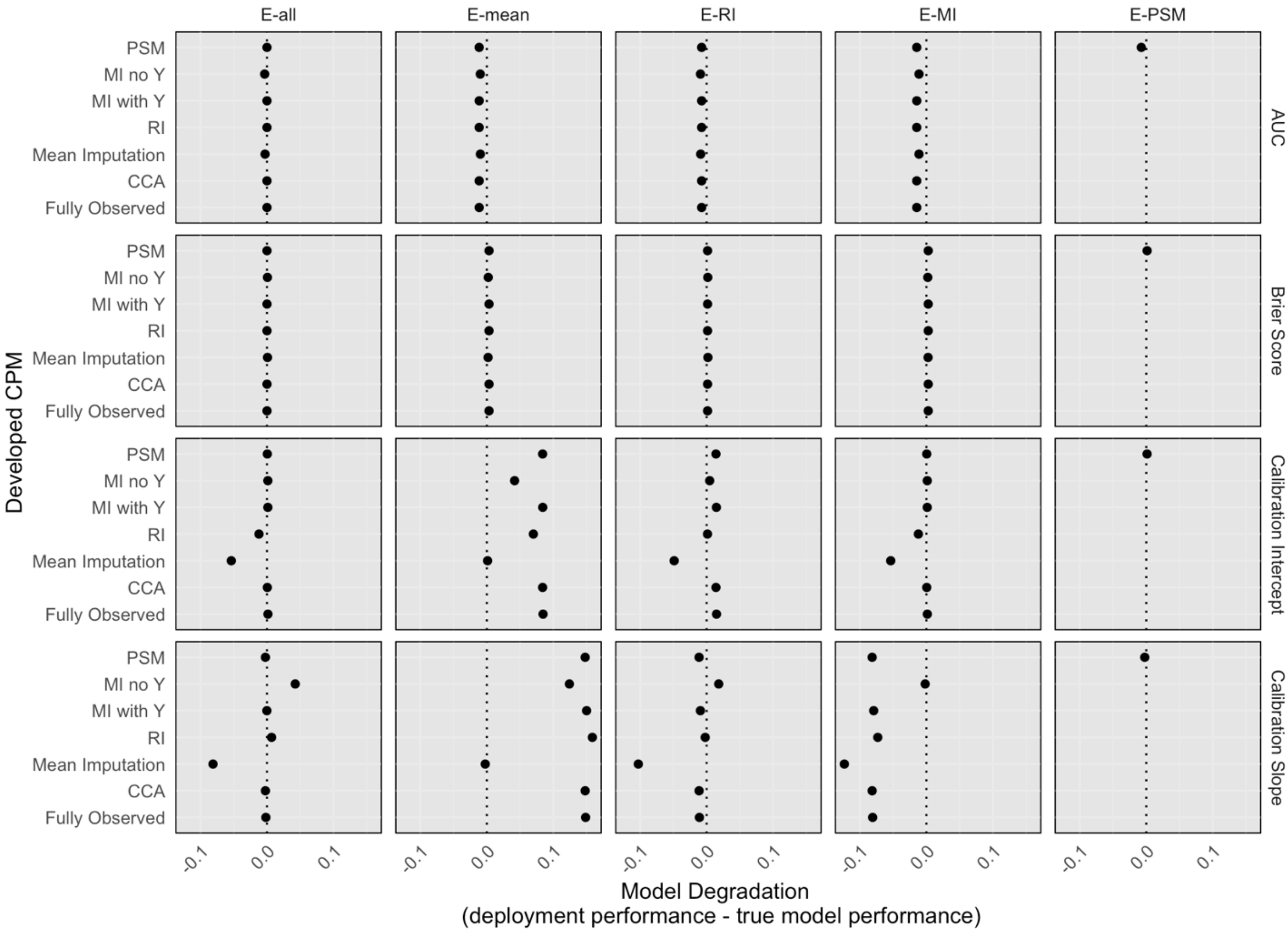

**Supplementary Figure 1:** Model Degradation of each developed CPM for each estimand, for scenarios were $X_1$ is continuous, contains 50% missing data, where $\gamma_1 = \gamma_2 = 0.5$, where the correlation between $X_1$ and $X_2$ ($\rho$) was 0.75, and under a consistent DAG(b) missingness mechanisms in development and validation sets. Model degradation was defined as the difference between the CPM's deployment performance for a given estimand, and the true performance for that estimand.

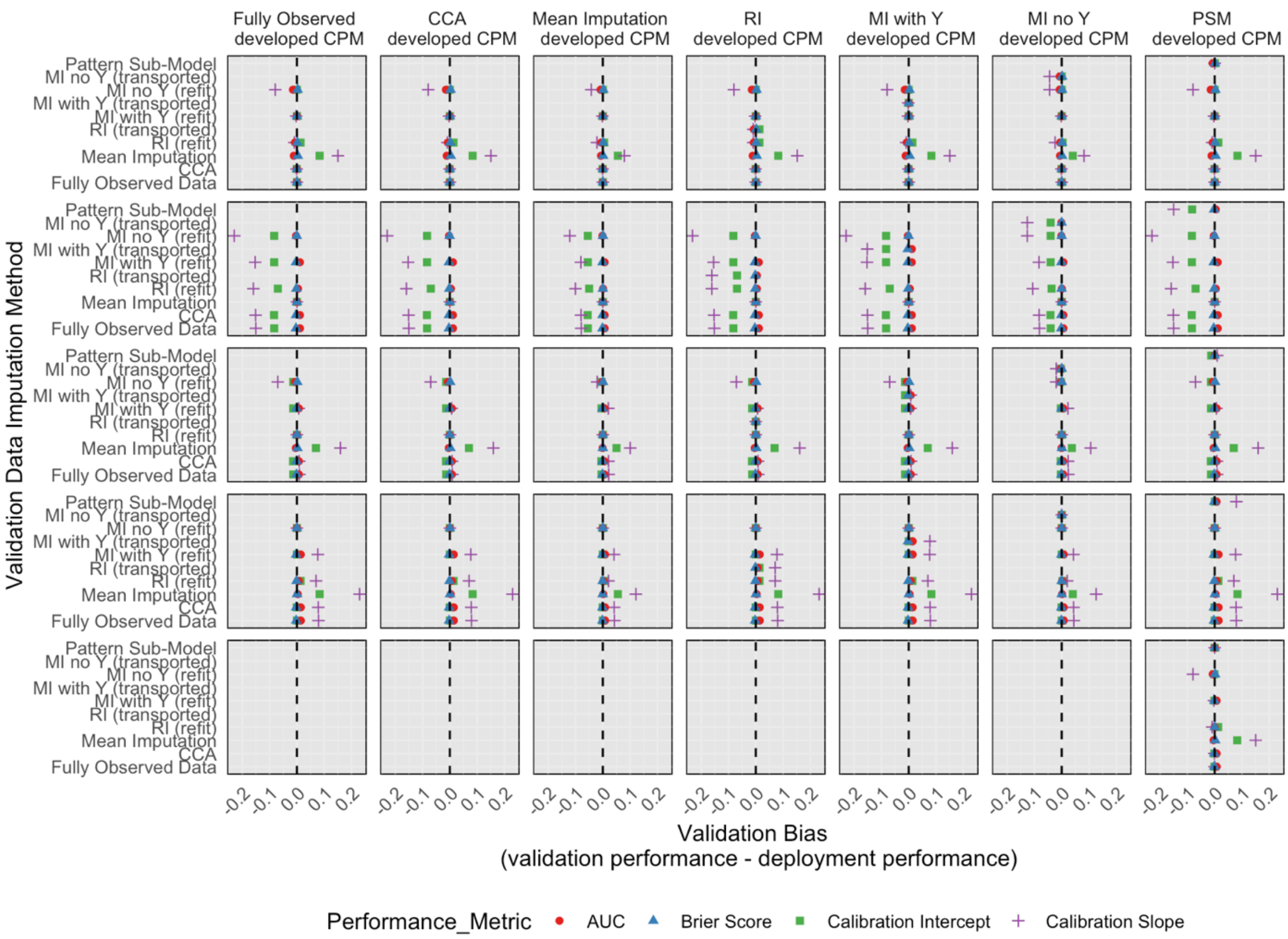

**Supplementary Figure 2:** Validation Bias of each developed CPM for each estimand, for scenarios were $X_1$ is continuous, contains 50% missing data, where $\gamma_1 = \gamma_2 = 0.5$, where the correlation between $X_1$ and $X_2$ ($\rho$) was 0.75, and under a consistent DAG(b) missingness mechanisms in development and validation sets. Validation bias was defined as the difference between a CPM's deployment performance and its validated performance under all alternative missing data handling strategies.

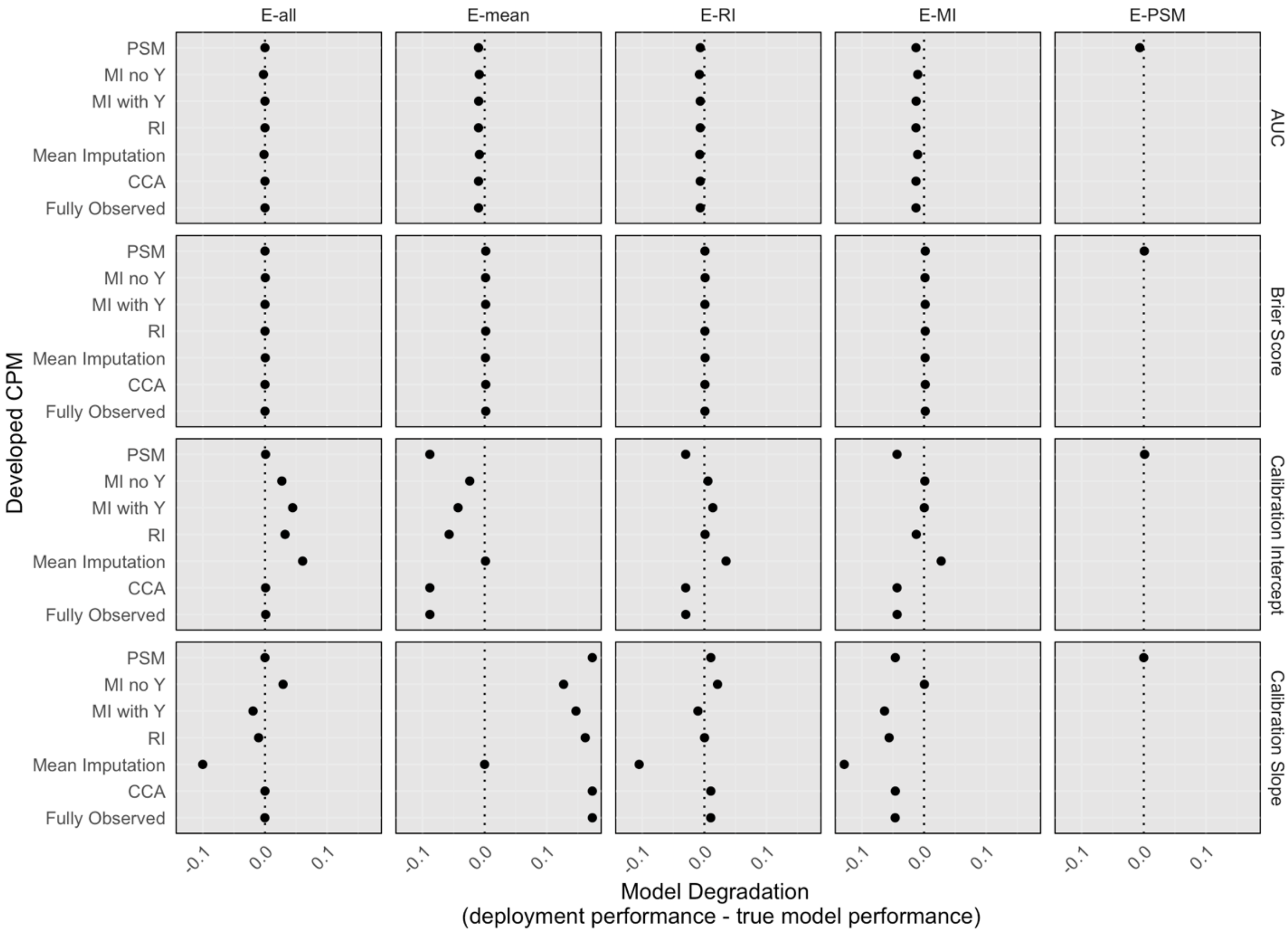

**Supplementary Figure 3:** Model Degradation of each developed CPM for each estimand, for scenarios were $X_1$ is continuous, contains 50% missing data, where $\gamma_1 = \gamma_2 = 0.5$, where the correlation between $X_1$ and $X_2$ ($\rho$) was 0.75, and under a consistent DAG(d) missingness mechanisms in development and validation sets. Model degradation was defined as the difference between the CPM's deployment performance for a given estimand, and the true performance for that estimand.

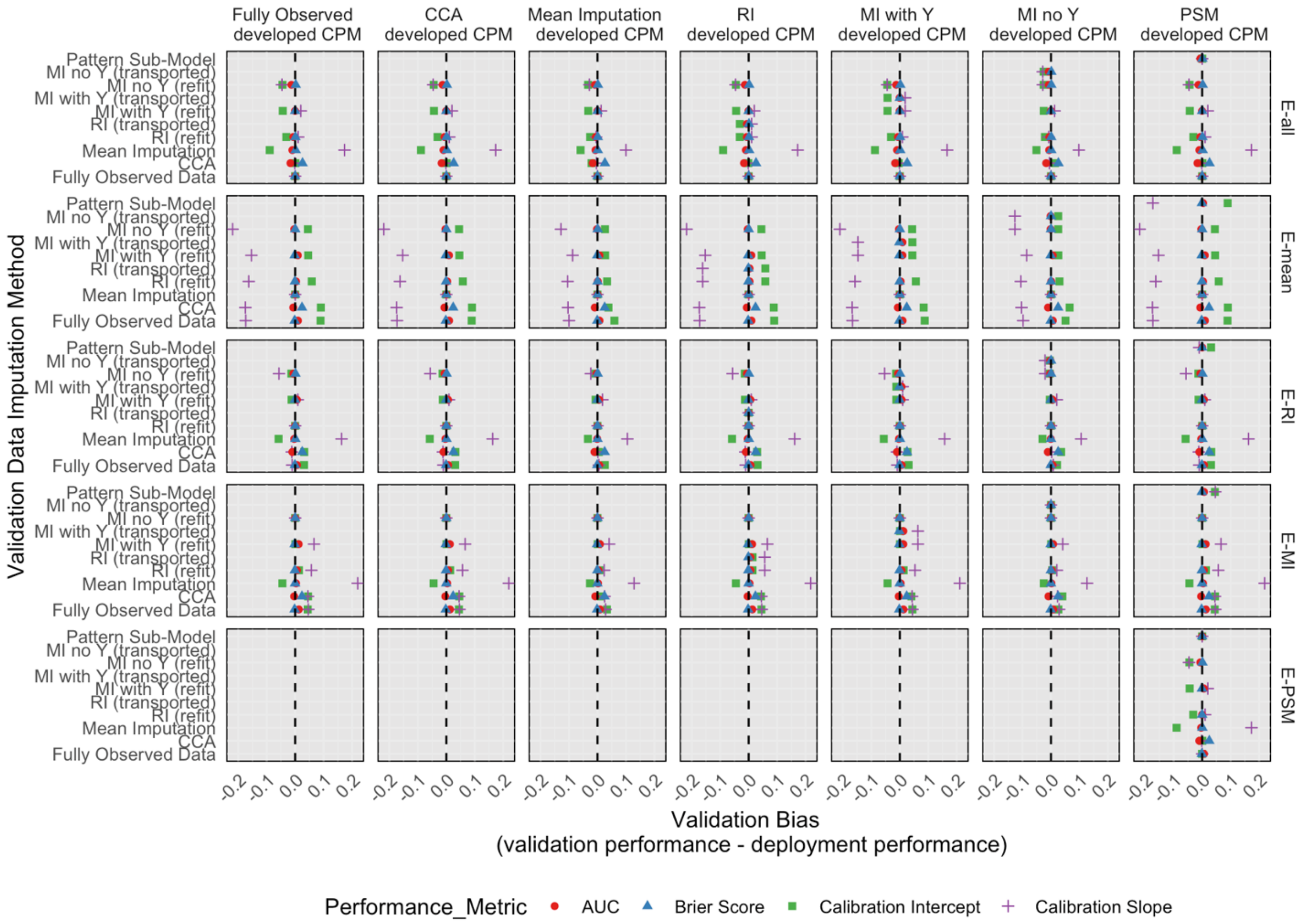

**Supplementary Figure 4:** Validation Bias of each developed CPM for each estimand, for scenarios were $X_1$ is continuous, contains 50% missing data, where $\gamma_1 = \gamma_2 = 0.5$, where the correlation between $X_1$ and $X_2$ ($\rho$) was 0.75, and under a consistent DAG(d) missingness mechanisms in development and validation sets. Validation bias was defined as the difference between a CPM's deployment performance and its validated performance under all alternative missing data handling strategies.

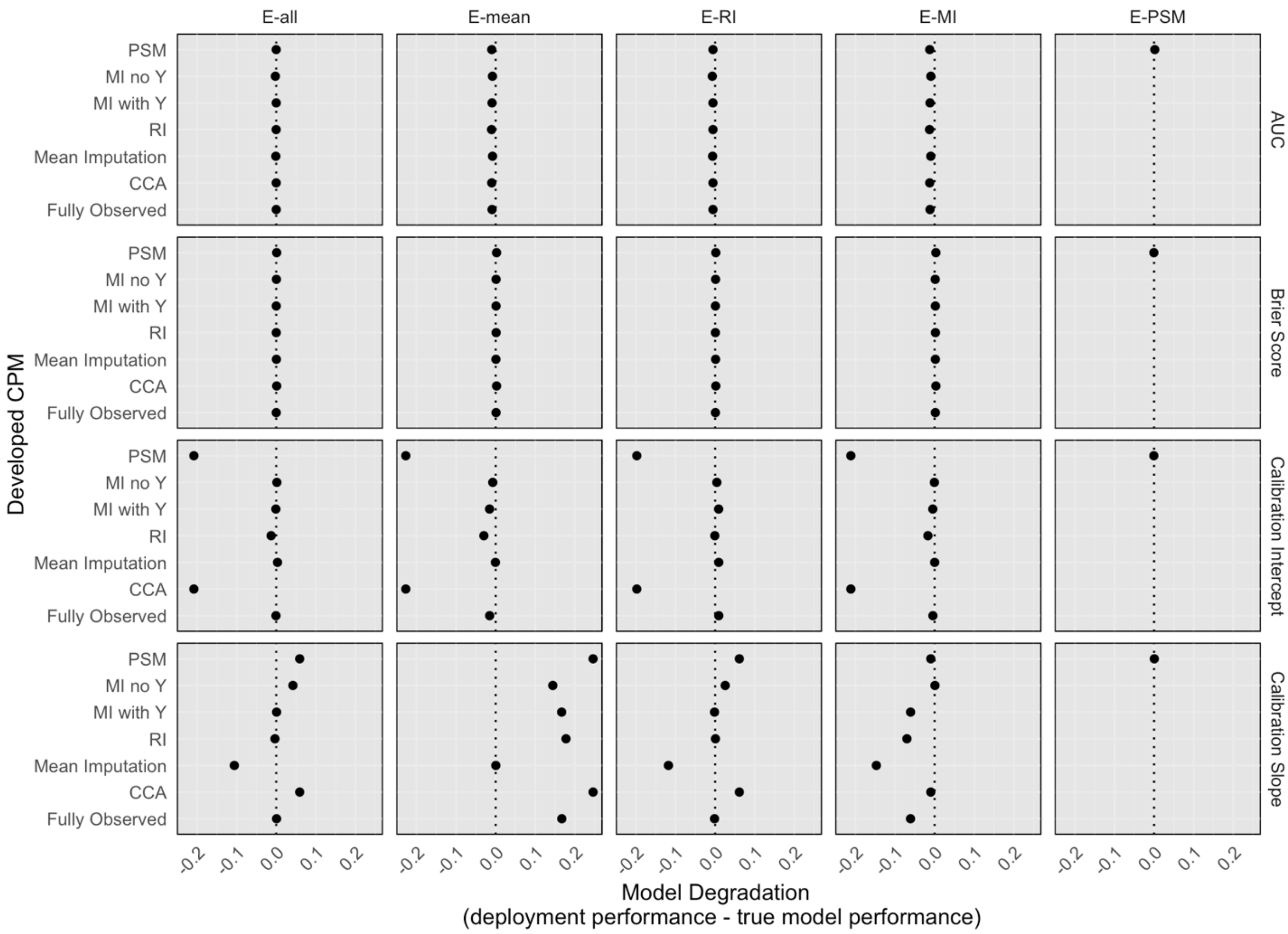

**Supplementary Figure 5:** Model Degradation of each developed CPM for each estimand, for scenarios were $X_1$ is continuous, contains 50% missing data, where $\gamma_1 = \gamma_2 = 0.5$, where the correlation between $X_1$ and $X_2$ ($\rho$) was 0.75, and under a consistent DAG(e) missingness mechanisms in development and validation sets. Model degradation was defined as the difference between the CPM's deployment performance for a given estimand, and the true performance for that estimand.

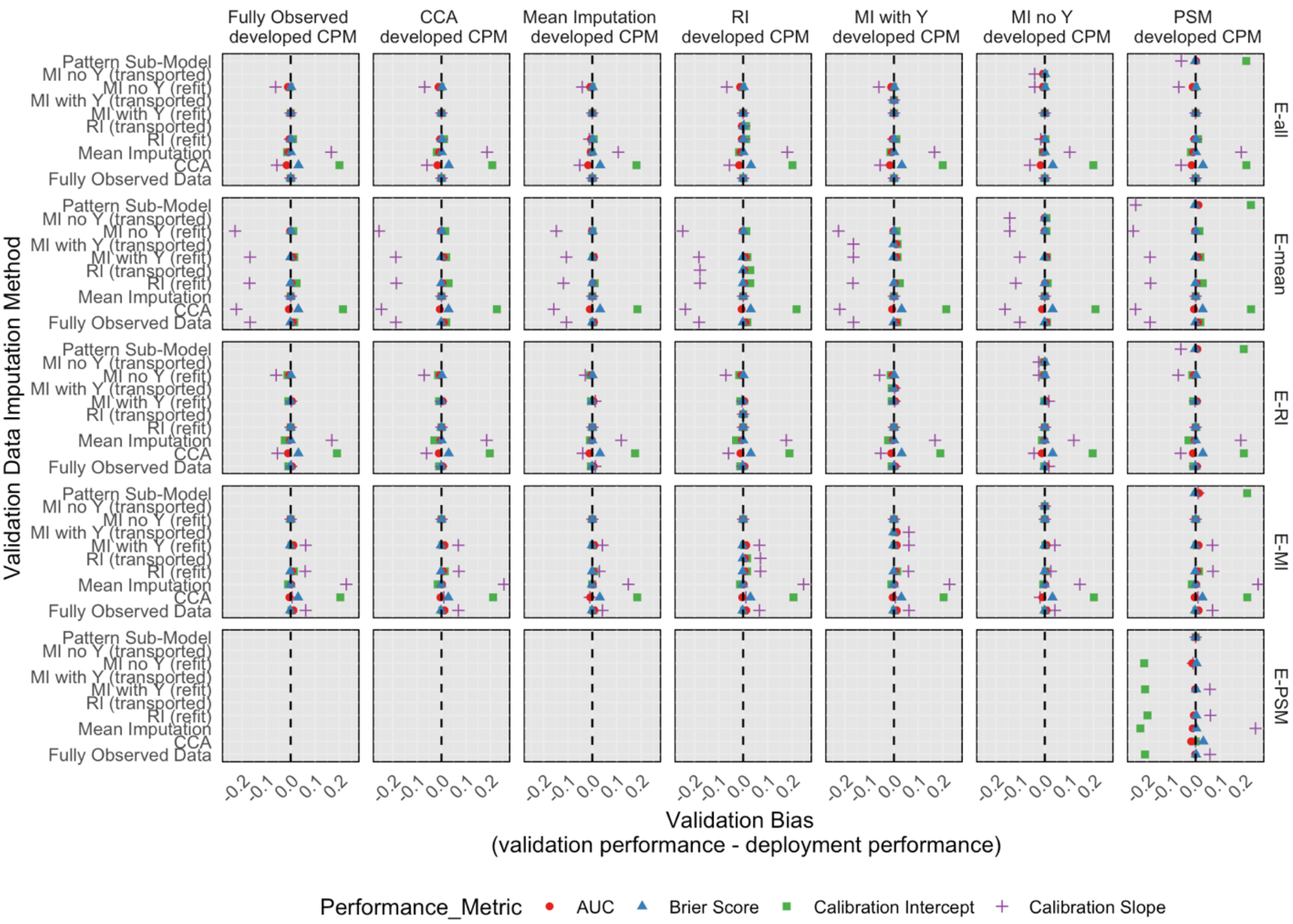

**Supplementary Figure 6:** Validation Bias of each developed CPM for each estimand, for scenarios were $X_1$ is continuous, contains 50% missing data, where $\gamma_1 = \gamma_2 = 0.5$, where the correlation between $X_1$ and $X_2$ ($\rho$) was 0.75, and under a consistent DAG(e) missingness mechanisms in development and validation sets. Validation bias was defined as the difference between a CPM's deployment performance and its validated performance under all alternative missing data handling strategies.